\documentclass[12pt]{article}
\usepackage{a4}

\begin{document}

\begin{center}

{\Large\bf On Invariants of Quark and Lepton Mass Matrices in the Standard Model}

\vspace{.3cm}
{\bf Cecilia Jarlskog}\\

Division of Mathematical Physics, Physics Department, LTH,\\
Lund University, Lund, Sweden
\end{center}

\vspace{.1cm}
\noindent
\begin{abstract}
\noindent
In this article I present the motivation for introducing the 
invariant functions of mass matrices, based on my own work, and give some examples.
Since their introduction in 1985, in the framework of the standard electroweak model, they have been 
used by many authors. Some authors have gone further along this path and have studied the 
extensions of this concept to frameworks beyond the standard model. I hope, in the near future, to
give a more detailed account of this subject, including recent developments.
\end{abstract}
\section{Introduction}
Since many decades the ``flavor puzzle'' has been on the minds of
many physicists. However, through the course of history, the 
language has changed and the emphasis has shifted. Nonetheless, the basic issues keep on
returning as rephrased questions.

\vspace*{.1cm}
\noindent
In 1983 I had the privilege of attending the annual meeting of the 
Norwegian Physical Society. One of the plenary speakers was the 
great Dutch physicist Hendrik Casimir (1909-2000).
His name is familiar to most physicists from the ``Casimir effect''. In group theory and particle physics
the ``Casimir invariants'' play a central role. Casimir, as a young man, had been at centers of 
``action'' in theoretical physics, such as with Niels Bohr in Copenhagen, and 
with Wolfgang Pauli, as his assistant, in Z\"{u}rich.
Later on, in 1946, he had left physics and gone to industry.
Therefore, it was particularly interesting to hear what this powerful voice from the past had to say.
The most surprising statement he made was that the
greatest puzzle in physics is why the ratio of masses of the proton and the electron
is $1836$! Casimir was thus telling us that already in the ``old days'', i.e., before the second world war,
the physicists had been concerned with what we now call the flavor problem, albeit in a much simplified version.
Richard Feynman (1918-1988) was also puzzled by the flavor problem. He considered the question 
``why the muon weighs'' to be
one of the most important ones in physics. By now, every particle physicist has heard of the famous 
statement ``who ordered that?'' by Isidor Rabi (1898-1988) when the muon was discovered.
The flavor problem has always been considered to be ``super-important''. 
But in spite of a huge amount of collected knowledge about this matter, the basic questions have
not yet been answered.

\section{Models of mass matrices}

The question of quark masses and mixings plays a leading role
in the flavor puzzle. During several decades theorists have been inventing 
models for mass matrices hoping to 
gain an insight into the ``underlying principles''.
The advent of grand unified theories, especially the Georgi-Glashow model, 
made it plausible that there may be a connection between the lepton and quark masses.
Since then many models have been proposed. In order to make the mass matrices as
simple and predictive as possible, many authors have tried
to put in as many zeros as possible into them. These zeros give
what is called the ``texture'' of the mass matrices. These textures have been studied and even tabulated. 
However, it is always good to have some measure of reliability of a proposed model.
In the case of textures an obvious question is: 
what is the significance of these zeros? One can make
a simple analogy using the theory of special relativity. Consider, for example,
the collision of two particles A and B. In the rest frame of B, the momentum of B is 
zero while that of A is nonzero.
By going to the rest frame of A, the zero in
the momentum of B evaporates and moves to the momentum of A. Furthermore,
in their center of mass system both zeros evaporate. Obviously, there is nothing
special about a zero in this case. It has no deeper ``meaning''. The meaningful quantities are kinematical
invariants such as $s = (p_{A}+p_{B})^{2}$, where the $p$'s stand for four-momenta. The masses are, of course,
frame-independent quantities, $M_{A}^{2} = p_{A}^{2}$, etc. Similarly, in the case of quark 
and lepton masses and mixings
there are frame-dependent and frame-independent quantities, but now in the flavor space.
Only frame-independent results can be trusted.

\vspace{.1cm}
\noindent 
In the case of special relativity we have learned what is meant by (inertial) frames 
and know the transformation rules. But what are the frames in flavor physics?
The answer is that these frames depend 
on the model used and the invariants within it.
Let us now turn to the case of quarks in the standard model.

\section{The invariants in the standard model}
In this short article, all I wish to remind the reader about the standard model
with $n$ families of quarks and
leptons, is that the Higgs sector produces what is called mass matrices for the up-type and
down-type quarks as well as for leptons. These are all n-by-n matrices.
For the neutrinos there is an additional ``complication'' because the right-handed
neutrinos are gauge-singlets and therefore can couple to themselves. Thus, there could be 
additional (Majorana) mass terms in the theory. 
However, within the framework of the standard model,
these terms are not considered to be ``nice'',
because one would like all masses
to originate from the phenomenon of spontaneous symmetry breaking.

\vspace{.1cm}
\noindent
Let us consider the case of the quarks and denote the mass matrices generated for the
up-type and down-type quarks
by $M_{u}$ and $M_{d}$ respectively. In the standard model with $n$ families these are
general n-by-n matrices that need not even be Hermitian.

\vspace{.1cm}
\noindent
The crucial observation that leads to the concept of invariant functions of mass matrices is 
that these matrices are not uniquely defined but are frame-dependent in the sense that
given any such pair
$M_u$ and $M_d$, one can obtain an infinite number of other equivalent pairs by
unitary rotations in the flavor space. 
The measurable quantities must be ``invariant functions'' under such rotations. 
These invariants were first introduced in \cite{ceja85a} and studied in more 
detail in \cite{ceja87}. 
Actually, what enters, in the standard model, is the pair
\begin{equation} 
S_u \equiv M_u M_u^\dagger~,~S_d \equiv M_d M_d^\dagger~.
\label{quarks} 
\end{equation}
In the frame $X$ these matrices are replaced by
$ X S_{u} X^{\dagger}$ and $X S_{d} X^{\dagger}$, 
where $X$ is unitary.
Thus the invariant functions of mass matrices,
$f (S_{u}, S_{d})$, satisfy the condition
\begin{equation}
f (S_{u}, S_{d}) =  f (XS_{u}X^{\dagger}, X S_{d}X^{\dagger})~.
\end{equation}
Trivial examples of such invariant functions are chains of powers of the mass matrices, i.e.,
$ tr (S_{u}^{i} S_{d}^{j}S_{u}^{k} S_{d}^{l} ...)$.
As is well known, one needs to diagonalize the mass matrices in order to identify the physical 
states. Let us consider the case that nature seems to have chosen, i.e., $n=3$.
First we note that there are two "extreme frames", 
one in which the up-type quark mass matrix is diagonal, i.e., 
\begin{equation} 
S_u =
\left( \begin{array}{ccc}
m^2_u & 0 & 0 \\
0 & m^2_c & 0 \\
0 & 0 & m^2_t
\end{array}
\right), ~~~~~ S_d = V
\left( \begin{array}{ccc}
m^2_d & 0 & 0 \\
0 & m^2_s & 0 \\
0 & 0 & m^2_b
\end{array}
\right) V^\dagger 
\end{equation}
where the $m$'s refer to the quark masses and $V$ is the quark mixing matrix. 
The other extreme frame is one in which the down-type quark mass matrix is 
diagonal, i.e.,
\begin{equation}
S_d = 
\left( \begin{array}{ccc}
m^2_d & 0 & 0 \\
0 & m^2_s & 0 \\
0 & 0 & m^2_b
\end{array}
\right), ~~~~~   
S_u = V^\dagger
\left( \begin{array}{ccc}
m^2_u & 0 & 0 \\
0 & m^2_c & 0 \\
0 & 0 & m^2_t
\end{array}
\right) V~.
\end{equation}
These constructions are analogs of going from one extreme kinematical frame, where
particle B is at rest to the other when the particle A is at rest. 
The reader may wonder why should one care about the invariant functions of mass matrices.
The original reason was that in 1980's one was looking for a measure of CP violation in
the standard model. The question asked was: could CP be maximally violated in the quark sector of the
standard model? After all parity is maximally violated in interactions mediated by the $W$-bosons.
As is often the case, there were conflicting opinions 
on what was meant by maximal CP violation. Some authors were advocating that CP
is maximally violated if the
CP phase in the quark mixing matrix is 90 degrees.
However, such a definition makes no sense because
there is no such unique CP phase. This phase is convention dependent:
your CP phase is in general a function of my CP phase and mixing angles.  
The point raised by the present author was
that such a measure can only make sense if it is frame-independent, i.e., it has to
be an invariant function of the quark mass matrices.

\section{The invariant function for CP violation with three families}
There is a unique invariant for CP violation in the standard model with three families \cite{ceja85a}.
It is given by the determinant of 
the commutator of the quark mass
matrices, $det [S_{u}, S_{d}]$,
\begin{equation}
det \left[ S_u,S_d \right]  = 2i J . v(S_u). v(S_d)
\label{detquark}
\end{equation}
\noindent where J is an invariant whose magnitude equals  twice the area of
any of the six by now well-known unitarity triangles \cite{stora88}. The quantities
$v(S_u) $ and $v(S_d) $ are (Vandermonde determinants) given by
\begin{equation} 
v(S_u) = (m^2_u-m^2_c)(m^2_c -m^2_t)(m^2_t -m^2_u)
\end{equation}
\begin{equation}
v(S_d)= (m^2_d-m^2_s)(m^2_s -m^2_b)(m^2_b -m^2_d) ~.
\end{equation}
Looking into literature, we see 
the determinant in Eq.(\ref{detquark}) 
in essentially every computation involving CP
violation, in the three-family version of the standard model. It appears, in all its glory, 
when all the six quarks enter on equal footing
but otherwise in a well-defined truncated form, 
where some factors are missing due to
assumptions made in the calculation \cite{ceja87}.
Examples of the first kind are the renormalization
of the $\theta$-parameter of QCD by the electroweak interactions and 
the calculation of the
baryon asymmetry of the universe in the standard model. An example of
the second kind is the computation of
the electric dipole moment of a quark, say the
down quark. Since in such a computation, the down quark appears in the external legs,
it is tacitly assumed
that we know the identity of this quark, i.e., $m_d \neq m_s$ and $m_d \neq m_b$.
Therefore the factors
$(m^2_d-m^2_s)$ and $(m^2_b -m^2_d)$ are missing but all
the other factors are present.

\vspace{.1cm}
\noindent
It should also be mentioned that
the absolute values of the elements of the quark mixing matrix
are measurable quantities and thus can be expressed as invariant functions. These functions
were constructed in \cite{ceja87} (see also \cite{cejaproj}).

\vspace{.1cm}
\noindent
The above commutator is the simplest in a family of commutators of 
functions of mass matrices (see the first paper in \cite{ceja85a}),
\begin{equation}
C(f,g) \equiv \left[ f(S_u),g(S_d ) \right]
\label{commutator}
\end{equation}
\noindent $f$ and $g$ being functions that are diagonalized with 
the same unitary matrices that
diagonalize $S_u$ and $S_d$ respectively. The determinants of these
commutators, which are also invariant functions, are given by 
\begin{equation}
det \left[ f,g \right] = 2i J. v(f). v(g) 
\label{fgdet}
\end{equation}
\noindent where $v(f) \equiv v(f(S_u))$, and $v(g) \equiv v(g(S_d))$. 
More explicitly
\begin{equation}
v(f) = \sum_{i,j,k}^{} \epsilon_{ijk} f_j f^2_k~= 
(f_1-f_2)(f_2-f_3)(f_3-f_1)~.
\label{vander}
\end{equation} 
\noindent The $f_j$ denote the three eigenvalues of the matrix 
$f(S_u)$ and the quantities 
related to the down-type quarks are defined similarly.
An essential point is that Eq.(\ref{fgdet}) holds irrespectively of whether $f$ and $g$ 
are hermitian or not. This property makes the above formalism applicable to 
neutrino oscillations.

\section{Other applications}
The amount of space allocated to this paper allows me only to quote a few other results. 
Similar invariants enter when one studies {\bf CP violation in the neutrino sector}
of the standard model \cite{cejanu}. For this case, the neutrino Majorana mass matrix is
largely irrelevant and we may introduce the analogs of the pair in Eq. (\ref{quarks}) for the
leptons,
$$ S_\nu \equiv M_\nu M_\nu^\dagger~,~S_l \equiv M_l M_l^\dagger$$
and the commutators
\begin{equation} 
\Delta^\pm \equiv \left[ e^{\pm 2i\xi S_\nu}~,~ S_l
\right]
\label{vaccom}
\end{equation}  
\noindent where $\xi$ is a kinematical parameter (related to oscillation length).
The determinant of these commutators are invariant functions of
lepton mass matrices related to CP violation in the leptonic sector. Using Eq.(\ref{fgdet}) we have  
\begin{equation}
det\Delta^\pm= 2i~J_{\nu}.v(S_l) . v(e^{\pm 2i\xi S_\nu })
\label{vacdet} ~.
\end{equation} 
\noindent Here $J_{\nu}$ is the leptonic analog of the CP invariant of the quark mixing matrix. 
Here it is more convenient to use the index $\nu$ instead of "lep" (for leptons) 
because when dealing with oscillations in matter the notation is
easily generalized. $J^\prime_{\nu}$ and $J^\prime_{{\bar \nu}}$ will then denote
the corresponding quantities for neutrino and antineutrino oscillations in matter.    
Furthermore, just as in the case of the quarks, $J_{\nu}$ is simply twice the area of 
any of the six leptonic unitarity
triangles. The two $v$'s are Vandermonde determinants. Because of lack of space, I don't give the
details here. It is indeed (at least intellectually) gratifying that such invariants can 
be constructed. One can also construct the corresponding invariants for the case of neutrino 
oscillations in matter (see \cite{cejanu}).

\vspace{.1cm}
\noindent
As a final application, I would like to mention the question of textures.
Theorists love zeros in mass matrices as they
make life simpler. However, we have seen that such zeros are not invariants!
Also the so called the quark-lepton complementarity relations
suffer from not being relations among invariants and are thus ill-defined.
Some aspects of this matter has been discussed in \cite{cejaqlc}.

\section{Outlook}
The issue of masses and mixings remains an unsolved problem
that deserves our attention. However, we should always keep
in mind that important results can't be frame-dependent.
So, if you have an important message to transmit, you should
be able to formulate it in an invariant form.

\section{Acknowledgment}
I wish to thank Jean-Marie Fr\`{e}re and Achille Stocchi for inviting me to write this article,
for publication in Comptes Rendus Physique (Acad\'{e}mie des sciences).

\end{document}